\documentclass[%
 reprint,
 amsmath,amssymb,
 aps,
nofootinbib
]{revtex4-2}

\usepackage{graphicx}
\usepackage{dcolumn}
\usepackage{bm}
\usepackage{hyperref}
\usepackage{enumitem}
\usepackage{times}
\usepackage[normalem]{ulem}

\usepackage{xcolor}

\begin{document}

\preprint{APS/123-QED}

\title{Cosmology from HSC Y1 Weak Lensing with Combined Higher-Order Statistics and Simulation-based Inference}

\author{Camila P. Novaes$^{1,2, 3}$}\email{camila.novaes@inpe.br}
\author{Leander Thiele$^{2,3}$}\email{leander.thiele@ipmu.jp}
\author{Joaquin Armijo$^{2,3}$}
\author{Sihao Cheng$^{4,5}$}
\author{Jessica A. Cowell$^{2,3,6}$}
\author{Gabriela A. Marques$^{7,8}$}
\author{Elisa G. M. Ferreira$^{2,3}$}
\author{Masato Shirasaki$^{9,10}$}
\author{Ken Osato$^{11,12,2}$}
\author{Jia Liu$^{2,3}$}
\affiliation{%
 $^{1}$Instituto Nacional de Pesquisas Espaciais, Av. dos Astronautas 1758, Jardim da Granja, S\~ao Jos\'e dos Campos, SP, Brazil}
 \affiliation{%
 $^{2}$Kavli IPMU (WPI), UTIAS, The University of Tokyo, 5-1-5 Kashiwanoha, Kashiwa, Chiba 277-8583, Japan}
  \affiliation{%
 $^{3}$Center for Data-Driven Discovery, Kavli IPMU (WPI), UTIAS, The University of Tokyo, Kashiwa, Chiba 277-8583, Japan}
 \affiliation{%
 $^{4}$Institute for Advanced Study, 1 Einstein Dr., Princeton, NJ 08540, USA}
 \affiliation{%
 $^{5}$Perimeter Institute for Theoretical Physics, 31 Caroline St N, Waterloo, ON N2L 2Y5, Canada}
  \affiliation{%
$^{6}$Department of Physics, University of Oxford, Denys Wilkinson Building, Keble Road, Oxford OX1 3RH, United Kingdom}
\affiliation{%
 $^{7}$Fermi National Accelerator Laboratory, P. O. Box 500, Batavia, IL 60510, USA}
 \affiliation{%
 $^{8}$Kavli Institute for Cosmological Physics, University of Chicago, Chicago, IL 60637, USA}
 \affiliation{%
 $^{9}$National Astronomical Observatory of Japan (NAOJ), National Institutes of Natural Sciences, Osawa, Mitaka, Tokyo 181-8588, Japan}
 \affiliation{%
 $^{10}$The Institute of Statistical Mathematics, Tachikawa, Tokyo 190-8562, Japan}
 \affiliation{%
 $^{11}$Center for Frontier Science, Chiba University, 1-33 Yayoicho, Inage, Chiba 263-8522, Japan}
 \affiliation{%
 $^{12}$Department of Physics, Graduate School of Science, Chiba University, 1-33 Yayoicho, Inage, Chiba 263-8522, Japan}




\date{\today}

\begin{abstract}
We present cosmological constraints from weak lensing with the Subaru Hyper Suprime-Cam (HSC)  first-year (Y1) data, using a simulation-based inference (SBI) method. 
We explore the performance of a set of higher-order statistics (HOS) including the Minkowski functionals, counts of peaks and minima, and the probability distribution function and compare them to the traditional two-point statistics. 
The HOS, also known as non-Gaussian statistics, can extract additional non-Gaussian information that is inaccessible to the two-point statistics. We use a neural network to compress the summary statistics, followed by an SBI approach to infer the posterior distribution of the cosmological parameters. 
We apply cuts on angular scales and redshift bins to mitigate the impact of systematic effects. 
Combining two-point and non-Gaussian statistics, we obtain $S_8 \equiv \sigma_8 \sqrt{\Omega_m/0.3} = 0.804_{-0.040}^{+0.041}$ and $\Omega_m = 0.344_{-0.090}^{+0.083}$, similar to that from non-Gaussian statistics alone. These results are consistent with previous HSC analyses and Planck 2018 cosmology. 
Our constraints from non-Gaussian statistics are $\sim 25\%$ tighter in $S_8$ than two-point statistics, where the main improvement lies in $\Omega_m$, with $\sim 40$\% tighter error bar compared to using the angular power spectrum alone ($S_8 = 0.766_{-0.056}^{+0.054}$ and $\Omega_m = 0.365_{-0.141}^{+0.148}$).
We find that, among the non-Gaussian statistics we studied, the Minkowski functionals are the primary driver for this improvement. 
Our analyses confirm the SBI as a powerful approach for cosmological constraints, avoiding any assumptions about the functional form of the data's likelihood. 
\end{abstract}

\maketitle



\section{\label{sec:introduction}INTRODUCTION}

There is no doubt about the importance of surveying the large-scale structure of the Universe and mapping the distribution of baryonic matter, as measurements of its clustering serve as a powerful cosmological probe. 
The matter clustering gives rise to the weak lensing signal, which can be measured as a distortion of galaxy shape and provide a direct measurement of the total matter density contrast. 
In fact, valuable cosmological constraints have been possible through analyses of lensing shear catalogs provided by Stage-III \footnote{Definition introduced by the Dark Energy Task Force report \citep{albrecht2006report}.} surveys, such as the Kilo-Degree Survey (KiDS) \cite{2015/kids}, the Dark Energy Survey (DES) \cite{2021/des} and the Subaru Hyper Suprime-Cam (HSC) \cite{2018/hsc}.
Still, with the next generation of experiments, including Euclid \cite{2022/scaramella}, and Vera Rubin Observatory Legacy Survey of Space and Time (LSST) \cite{2012/lsst},  we expect to be able to map the distribution of matter deeper in the Universe, with higher source density and better shape measurements. 
This would allow us to investigate tensions in the current cosmology scenario, in particular the apparent discrepancy identified between the growth-of-structure parameter, $S_8$, from CMB measurements, as well as CMB lensing, and weak lensing data \cite{2021/di-valentino}.

Unlike the primary CMB temperature fluctuations, a nearly Gaussian field with imprints of the primordial density perturbations, weak lensing fields carry non-Gaussian (NG) information introduced by the non-linear evolution of cosmic structures. 
As a consequence, the traditional two-point statistics are insufficient to capture all the cosmological information contained in the data. 
To maximize the amount of information extracted from such datasets, it is crucial to make use of NG statistics, also known as higher-order statistics (HOS). 
With such motivation, several studies have been employing NG statistics for weak lensing probes \cite{2024/gatti-des3-I, 2023/anbajagane, 2023/liu, 2019/marques}. 
In particular, HSC first-year (Y1) weak lensing data have been explored using peaks and minimum counts (PM)  \cite{2024/marques-hsc}, probability distribution function (PDF) \cite{2023/thiele-hsc}, scattering transform \cite{2024/cheng-hsc}, and Minkowski functionals (MFs) \cite{2024/armijo-hsc}, while the first three have also been studied in the context of baryonic feedback \cite{2024/grandon-hsc}.

In this work, we use the MFs, PM, and PDF of the Subaru Hyper Suprime-Cam (HSC) first-year (HSC Y1) data, and their combination with the angular power spectrum (PS), to constrain the cosmological parameters $\Omega_m$, the total matter density today, and $S_8 = \sigma_8\sqrt{\Omega_m/0.3}$, written as a function of the linear matter fluctuation within 8$h^{-1}$ Mpc radius, $\sigma_8$.
Since NG statistics, in general, have no analytical expression or a known closed-form theoretical model, it is necessary to rely on simulation-based models to describe a given summary statistic. 
By forward modelling the statistics allows, for example, to build emulators for cosmological constraints, as implemented by previous analyses using NG statistics of the HSC Y1 data \cite{2024/marques-hsc, 2023/thiele-hsc,2024/cheng-hsc,2024/grandon-hsc,2024/armijo-hsc}.
However, this type of analysis requires making approximations, as the assumption of a Gaussian likelihood for the data. 
Here, instead, we forward model the summary statistics of HSC Y1 and perform a simulation-based inference (SBI) \cite{2020/cranmer}, also known as likelihood-free or implicit-likelihood inference.
The SBI approach, then, might be considered as an optimal solution given that it does not require a functional form for the likelihood. 
A similar approach is presented in Refs \cite{2024/gatti-des3-I, 2023/lin, 2024/massara-simbig, 2024/von-wietersheim-kramsta-kids, 2021/jeffrey}, which uses different Gaussian and non-Gaussian statistics to extract the cosmological information, while Refs \cite{2023/lemos-simbig, 2022/zhao, 2021/jeffrey} rely on convolutional neural networks (CNN) for feature extraction (see Ref. \citep{2023/lu-hsc} for an application to HSC Y1).

The pipeline implemented for our analyses consists of two main steps:
\begin{itemize}[leftmargin=*]
    \item[1.] a neural network (NN) compresses the summary statistics (neural compression), or, in other words, it maps between summary statistics and cosmological parameters $(\Omega_m,S_8)$, providing the predicted (compressed) values for these parameters (following Ref. \cite{2024/novaes});
    \vspace{-0.2cm}
    \item[2.] a neural density estimator (NDE), trained over the compressed data (from simulations), infers the posterior distribution of $\Omega_m$ and $S_8$ parameters when evaluated on the measured data (from simulations or observation) after the same compression.
\end{itemize}    \vspace{-0.2cm}
We notice that data vectors comprising a summary statistic, or a combination of statistics, can be large enough to make the SBI process computationally prohibitively expensive. 
Also, SBI does not handle large data vectors well with a finite number of simulations \cite{2021/hermans}. 
This motivates the first step of our pipeline, including a massive compression of the summary statistics. 
Our methodology brings a novel approach in comparison to previous weak lensing NG analyses, showing that the combination of several summary statistics is promising and feasible in terms of SBI analyses. 
In particular, we demonstrate the feasibility of the SBI approach with a limited number of simulations, a common limitation in weak lensing analyses, given their computational cost.

The paper is organized as follows. 
In Section \ref{sec:data-simulations} we briefly describe the HSC Y1 data and simulations. 
Section \ref{sec:stats} presents the summary statistics employed in our analyses, while Section \ref{sec:sbi} presents our methodology for cosmological analyses, describing the NN algorithm employed for data compression and the likelihood-free approach for parameter inference. 
In Section \ref{sec:systematics} we show the robustness of our results against systematic effects. 
Our results are presented in Section \ref{sec:results}, and we conclude in Section \ref{sec:conclusion}.

\section{\label{sec:data-simulations}Data and simulations}

In this section, we briefly describe the HSC Y1 lensing convergence data, as well as the simulations employed in our analysis. 
We refer the reader to Ref. \cite{2024/marques-hsc} for additional information. 

\subsection{HSC Y1 data}

As a precursor to Stage IV surveys, HSC is a deep, large-scale lensing survey with a high density of source galaxies. Here, we utilize the shear catalog from the first public data release, HSC Y1 \cite{2018/mandelbaum-hsc-data}\footnote{Publicly available at \url{https://hsc-release.mtk.nao.ac.jp/doc/index.php/s16a-shape-catalog-pdr2/}}, selecting galaxies with reliable shape measurements covering 136.9 deg$^2$ of the sky in six disjoint patches and spanning a redshift range from 0.3 to 1.5. 
In our analysis, we define the redshift bins based on photometric redshifts determined using the {\tt MLZ} code \cite{2018/tanaka-hsc}. 
We split the source galaxies into four tomographic redshift bins, namely, $0.3 < z_1 < 0.6$, $0.6 < z_2 < 0.9$, $0.9 < z_3 < 1.2$, and $1.2 < z_4 < 1.5$, from which we use only the first three, given possible contamination by unknown systematics in the highest $z$-bin~\citep{2023/dalal-hsc3, 2023/thiele-hsc}. 
After creating pixelized maps from the shear catalog for each redshift bin and patch, they are smoothed with a Gaussian kernel and used to construct the corresponding convergence map using the Kaiser-Squires inversion method \cite{1993/kaiser-squires}. 
Our main analysis employs the smoothing scale\footnote{We use the Gaussian kernel defined as: $W_G = [1/ (\pi \theta^2_s)] \exp(-\theta^2/\theta^2_s)$.} of $\theta_s = 4$ arcmin, although we also evaluate our constraining power when adding $\theta_s = 8$ arcmin. 

\subsection{Simulations}

To forward model the summary statistics, we use the suite of N-body simulations introduced in Ref. \cite{2021/shirasaki-hsc}, produced for 100 different cosmologies sampled in the ranges $0.1 < \Omega_m < 0.7$ and $0.23 < S_8 < 1.1$, while fixing the Hubble parameter, baryon density, dark energy equation-of-state parameter, and spectral index, respectively, at $h = 0.6727$, $\Omega_b = 0.049$, $w = -1$, and $n_s = 0.9645$.
These simulations are based on the parallel Tree-Particle Mesh code \texttt{Gadget-2}, with $512^3$ particles. A typical value of the thickness of the mass sheet is about $150$, $200$, and $300\,h^{-1}$Mpc at $z<0.5$, $0.5<z<1$, and $1<z<2$, respectively. To create convergence maps through the ray-tracing technique, we adopt a flat-sky approximation and the multiple lens-plane algorithm \citep{Jain_2000,hamana_takashi}. 
For each pair of $(\Omega_m,S_8)$ values, a total of 50 realizations are constructed by randomly selecting observer positions in the simulation box, providing a set of quasi-independent realizations from only one simulation. This set of simulations is hereafter called \textit{cosmology-varied}. 

We also use a second set of simulations comprised by 2268 realizations generated from 108 quasi-independent full-sky simulations \cite{2017/takahashi} with cosmology fixed at the Wilkinson Microwave Anisotropy Probe nine-year data best-fit \cite{2013/hinshaw-wmap9}: $\Omega_b = 0.046$, $\Omega_m = 0.279$, $\Omega_\Lambda = 0.721$, $h = 0.7$, $S_8 = 0.79$, and $n_s = 0.97$. 
We refer to this set of simulations at a fixed cosmology as our \textit{fiducial} set. 

Both sets of simulations are tailored to HSC Y1 data, accounting for realistic aspects such as the survey geometry, multiplicative biases, redshift uncertainty, spatial inhomogeneities of source galaxies, shape noise levels, among other observational characteristics, following Refs \cite{2014/shirasaki-yoshida,2024/marques-hsc}. The simulations also account for the same smoothing scales applied to the data.

\section{\label{sec:stats}Summary statistics}

Our analyses assess the constraining power of one Gaussian statistic, the PS, and three NG statistics, the MFs, PM, and PDF. 
Below, we summarize the main details of their estimates from the convergence maps.

\vspace{0.1cm} 
\paragraph*{\bf Angular power spectrum.}
We estimate the PS using the pseudo-$C_\ell$ method implemented in the {\tt NaMaster} public code \cite{2019/alonso-namaster}\footnote{\href{https://namaster.readthedocs.io/en/latest/}{https://namaster.readthedocs.io/en/latest/}}, accounting for the geometry of each sky patch.
We estimate the  pseudo-$C_\ell$, $C_\ell^{\kappa\kappa}$, in 14 logarithmically spaced multipole bands in the total range of $81 < \ell <6500$. 
Out of this range, we use $300 < \ell < 1000$, i.e., 4 multipole bands in our analyses. 
Following Refs \cite{2024/marques-hsc,2023/thiele-hsc}, we consider the lower limit as established by Ref. \cite{2018/oguri-hsc}, according to assessments of unmodeled systematic errors on large scales. 
The maximum multipole is defined considering possible bias introduced by baryonic effects (see Ref. \cite{2024/grandon-hsc} for detailed analyses on the impact of these effects on Gaussian and non-Gaussian statistics). 
For a given redshift bin, a single $C_\ell^{\kappa\kappa}$ data vector is obtained by a weighted average over the six sky patches,  $C_\ell^{\kappa\kappa} =  \sum_i w_i C_\ell^{\kappa\kappa}|_i $, where $w_i$ is proportional to the inverse of the number of galaxies in the $i$th sky patch. \vspace{0.1cm}

\vspace{0.1cm} 
\paragraph*{\bf Minkowski functionals.}
Widely used in CMB \citep{2003/komatsu, 2013/modest, 2015/novaes, 2016/novaes, 2019/planckVII} and large scale structure analyses \cite{2007/saar,2010/kerscher,2013/choi, 2018/novaes}, the Minkowski functionals (MFs) \cite{1903/minkowski,1999/novikov} carry morphological information of structures, requiring $N+1$ functionals to completely characterize an N-dimensional field. 
Here, we use three MFs, the Area ($V_0$), Perimeter ($V_1$), and Genus ($V_2$), to access the NG features in the 2D lensing convergence maps, $\nu = \kappa/\sigma_0$, normalized by the average of the standard deviation $\sigma_0$ of the $\kappa$ fields composing the 2268 fiducial simulations. 
The three quantities are given by Ref. \cite{1999/novikov,2013/ducout}
\begin{align}
\! V_0(\nu_t) &= \frac{1}{4 \pi} \int_{\Sigma} d\Omega \, ,  \label{eq:v0}\\
\! V_1(\nu_t) &= \frac{1}{4 \pi} \frac{1}{4} \int_{\partial\Sigma} dl \, ,  \label{eq:v1}\\
\! V_2(\nu_t) &= \frac{1}{4 \pi} \frac{1}{2 \pi} \int_{\partial\Sigma} \zeta~dl  \label{eq:v2}\, ,
\end{align}
and calculated for excursion sets $\Sigma \equiv \{ \kappa > \nu_t \sigma_0 \}$, i.e., sets of connected pixels whose values exceed a $\nu_t$ threshold, with boundaries $\partial \Sigma \equiv \{ \kappa = \nu_t \sigma_0 \}$. 
The quantities $d\Omega$ and $dl$ are the solid angle and line elements, respectively, and $\zeta$ is the geodesic curvature. 
Here, we estimate the MFs for 19 $\nu_t$ thresholds linearly spaced in $[-4,4]$. 
More information on MFs and their application to lensing data can be found in, e.g., Refs \cite{2014/shirasaki-yoshida, 2015/petri, 2017/shirasaki, 2017/matilla, 2012/kratochvil, 2019/marques}. 
The MFs have also been used for HSC Y1 analyses by Ref. \cite{2024/armijo-hsc}. 
Note that, for each redshift bin, we obtain the total MFs by summing the results from each of the six HSC sky patches. 
The same procedure is employed for the other NG statistics. \vspace{0.1cm}

\vspace{0.1cm} 
\paragraph*{\bf Peaks and minima counts.}
Peaks counts have been widely explored for weak lensing analyses \cite{2015/liu,2015/liu-CFHTLenS} and, more recently, the minimum counts have also been shown to be efficient in cosmological constraints from weak lensing data \cite{2020/coulton}. 
A peak is defined as a pixel whose value $\nu \equiv \kappa / \sigma_0$ exceeds the values from the eight neighboring pixels, while a minimum is the opposite, exhibiting a lower value compared to the eight neighboring pixels. 
We count the peaks and minima (PM) within bins defined by the same 19 $\nu_t$ values as the MFs, 
following previous applications of these statistics to HSC Y1 data \cite{2024/marques-hsc,2024/grandon-hsc}. 
\vspace{0.1cm}

\vspace{0.1cm} 
\paragraph*{\bf Probability distribution function.}
Sensitive to NG signatures of a field, the probability distribution function (PDF) is also a promising summary statistic for cosmological constraints \cite{2019/liu}, as demonstrated by Ref. \cite{2023/thiele-hsc} from analysis of the HSC Y1 data. 
Here, similarly to MFs and PM, we estimate this one-point statistic as the histogram of the pixels in the normalized convergence maps, $\nu \equiv \kappa/\sigma_0$, for the same 19 $\nu_t$ bins as previously defined in this section.

\section{\label{sec:sbi}Simulation-based inference}

Our purpose for this paper is to infer cosmology from HSC Y1 data following a procedure fully based on simulations. 
We rely on the forward modelling of both Gaussian and non-Gaussian statistics without assuming a functional form for the likelihood (likelihood-free inference). 
The combination of the summary statistics considered here, estimated from three redshift bins, results in data vectors of length 354 (see discussion in the next subsection). 
The performance of a neural density estimator to infer the posterior distribution directly from such a large data vector is limited by the number of simulations, besides being highly computationally expensive. 
Therefore, our SBI pipeline includes a first step of data compression. 
This is intended to reduce the dimensionality of the data vector while keeping as much information as possible. 
The posterior estimation is then performed over the compressed data.

\subsection{Neural network and data compression}

The data compression can be performed following different approaches \cite{2000/heavens-moped,2021/jeffrey,2018/charnock}. 
The method adopted in this paper uses an NN (neural compression; examples can be found in Refs \cite{2024/gatti-des3-I,2024/jeffrey, 2021/jeffrey}). 
In this case, the NN works as a function $\mathcal{F}$ applied to the N-dimensional summary statistics $X$ and returns a compressed data vector 
\begin{equation}
    x = \mathcal{F}(X),
\end{equation}
which keeps the maximum information about the cosmological parameters of interest, $\theta$ \cite{2024/jeffrey}. 
The dimension of the compressed data can be chosen to match that of $\theta$. 
We notice that an alternative way of looking at this problem is to take the NN as a way of mapping the summary statistics, a set of features, to the cosmological parameters, taken as targets for the loss function, so that $x$ would be the prediction of the cosmological parameters \cite{2024/novaes, 2022/villaescusa, 2022/perez, 2015/novaes}. 

We summarize below our implementation of NNs for neural compression. 
We closely follow the methodology for the training process and the choice of hyperparameters as done in Ref. \cite{2024/novaes}, to which we refer the readers for further details. \vspace{0.1cm}

\paragraph*{\bf Training and test.}
We use the summary statistics estimated from the cosmology-varied set of simulations for training and testing the NN algorithm: 50 realizations for each of the 100 cosmologies, or pairs $\{\Omega_m,S_8\}$, sampling the parameter space.
The summary statistics, $X$, feed the NNs and the corresponding cosmological parameters are the outputs, $x$ (values to be predicted).  
Since we have a restricted number of cosmologies, we maximize the information fed into the NN during training by following the idea of ``leave-one-out" \cite{2023/geroldinger}.
This means that we train an NN 100 times, keeping fixed the hyperparameters, each time leaving one cosmology out. 
We use the corresponding set of 50 realizations as our test set.
In this way, we end up with 100 NN trained models, i.e., ${\mathcal F}_i$ for $i=1,...,100$, and each of them is tested over the summary statistics from the 50 realizations of the excluded cosmology, $i$, ensuring that none of the models is tested over cosmologies used for training. 
The 99 cosmologies used for each training are split so that 20\% of them is reserved for validation. 
Although, for simplicity, we refer to the summary statistics from the 99 cosmologies as the training set, a cross-validation procedure is always employed (see Ref. \cite{2024/novaes}). 

The training set is defined as $\mathcal{T}(X,\theta) = \mathcal{T}(X_j,\theta_j)$, allowing the NN to learn the relation between a given summary statistic (or combination of them), $X$, and the true cosmological parameters, $\theta$. 
Then, the $X_j$ data vector for the $j$th realization is given by 
\begin{equation}
    X_j = \{ (\mathtt{S})^z \}_j,
\end{equation}
where $\mathtt{S}$ is composed by one or more summary statistics and $z$ runs over the redshift bins, $z_1$, $z_2$, and $z_3$. 
For example, combining all 4 statistics, we have a data vector $(\mathtt{S}) = (C_\ell^{\kappa \kappa}, {\rm MFs}_\nu, {\rm PM}_\nu, {\rm PDF}_\nu)$ of length 118 for an individual redshift $z$, and length 354 when combining all three redshift bins. 
The training is performed for each cosmological parameter at a time, so that an NN learns the relation between $X$ and $\theta = \{ \Omega_m \}$, and  $X$ and $\theta = \{ S_8 \}$, separately. \vspace{0.1cm}

\paragraph*{\bf Architecture.}
We use a fully connected NN optimizing the set of hyperparameters for each individual test performed (e.g., different combinations of summary statistics or redshift bins). 
For this, we use the {\tt optuna} package \cite{2019/akiba_optuna}, which allows us to automatically sample the hyperparameter space, fixing the maximum number of trials (the number of tested architectures) to 600, although the optimized set is generally chosen between the 350 and 500 trials. 
We use the mean square error ({\sc mse}) as our loss function: 
\begin{equation} \label{eq:mse}
\mbox{\sc mse} = \frac{1}{N} \sum_{i=1}^{N} (x_i - \theta_i)^2,
\end{equation}
where $N$ denotes the number of simulations in the validation set and $x_i,\theta_i$ the prediction and true values of a cosmological parameter ($\Omega_m$ or $S_8$). 
Note that we use all the 100 cosmologies to search for the optimized set of hyperparameters.
After that, we fix the hyperparameters\footnote{By fixing the optimizer to {\tt Adam} with $\beta_1,\beta_2 = 0.9, 0.999$, {\tt optuna} searches for the main hyperparameters, with ranges/options for each of them as follows: 
1) number of neurons in the hidden layers: [1,500];
2) number of layers: [1,3];
3) activation function: {\tt ReLu}, {\tt tanh};
4) learning rate: [$10^{-4}$,$10^{-2}$];
5) batch size: [50,500]; and
6) number of epochs: [50,500]. 
The 600 trials for one individual test usually take no more than 24 hours in a single Intel(R) Xeon(R) Gold 6230R CPU @ 2.10GHz.} and train the NN 100 times, each of them excluding one of the cosmologies, obtaining 100 $\mathcal{F}_i$ models.

\begin{figure}[h]
\includegraphics[width=0.97\columnwidth]{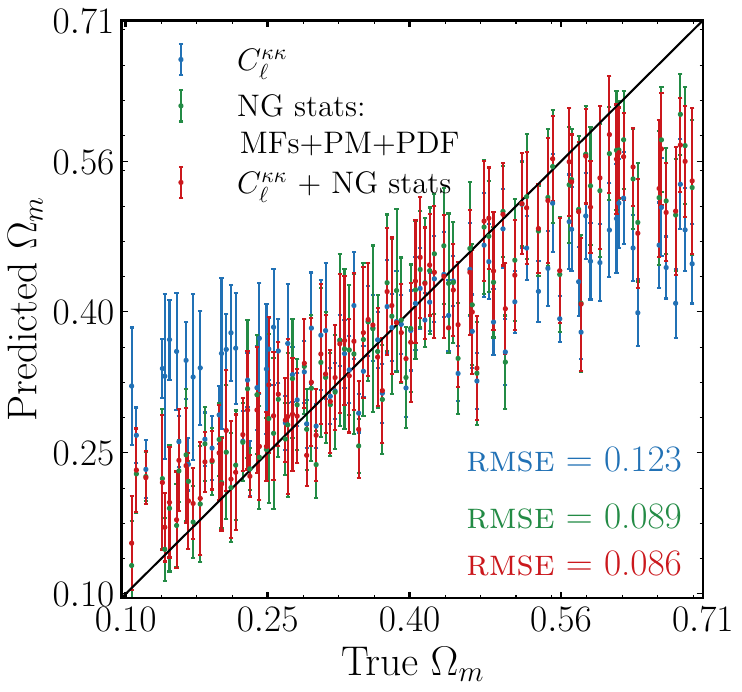}
\includegraphics[width=0.97\columnwidth]{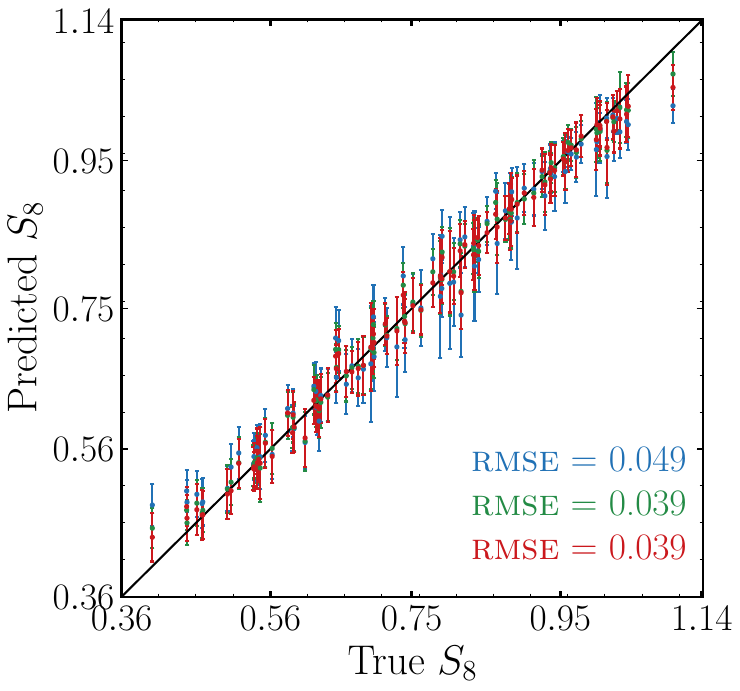}
\caption{\label{fig:NNpred} Scatter plots of the average of the 50 predicted (compressed) values of $\Omega_m$ (top) and $S_8$ (bottom) as a function of the true values of the cosmological parameters. The error bars correspond to the standard deviation of the respective set of 50 values. 
The different colors represent the different sets of summary statistics used for neural compression: the PS alone (blue), the combination of all three NG statistics (green), and the combination of the PS and NG statistics (red). Following the same color scale, we show in the bottom right of each panel the $\mbox{\sc rmse}$ value for each case, estimated from the 5000  predictions (100 cosmologies $\times$ 50 realizations) and the respective true values. The black diagonal line represents the identity $x=\theta$. }
\end{figure}

\paragraph*{\bf Compressed (predicted) data.}
Each NN model $\mathcal{F}_i$ obtained is then applied to the summary statistics estimated from the 50 realizations of $i$th cosmology excluded from the training, The average and standard deviation from the 50 predicted (compressed) values, $x$, for $i = 1, ..., 100$, are shown in Figure \ref{fig:NNpred} as a function of the true values, for both cosmological parameters. 
We summarize the results for three different combinations of summary statistics: all of them together, $(\mathtt{S}) = (C_\ell^{\kappa \kappa}, {\rm MFs}_\nu, {\rm PM}_\nu, {\rm PDF}_\nu)$, only the NG statistics, $(\mathtt{S}) = ({\rm MFs}_\nu, {\rm PM}_\nu, {\rm PDF}_\nu)$, and the PS alone $(\mathtt{S}) = (C_\ell^{\kappa \kappa})$. 
All of our tests combine the three redshift bins unless stated otherwise. 
Following the same color code, we show the root-mean-square error ($\mbox{\sc rmse} = \sqrt{\mbox{\sc mse}}$, with $\mbox{\sc mse}$ given by Eq. \ref{eq:mse}), quantifying the error (accuracy) of the NN predictions. 
These values show that combining Gaussian and NG statistics allows $\Omega_m$ and $S_8$ parameter predictions to improve by 30\% and 21\%, respectively, when compared to using PS only. 
The tighter scatter of $S_8$ predictions, compared to $\Omega_m$, indicates that the data is more informative about $S_8$ parameter relative to the prior range. 
The predictions for $\Omega_m$ are poorer than for $S_8$, as also found from a similar analysis of DES data, as discussed in Ref. \cite{2024/gatti-des3-II}. 
In other words, the deviation from the identity, represented by the black solid lines in Figure \ref{fig:NNpred}, happens when the features (summary statistics) have not enough information to allow the NN to efficiently map them in terms of the targets (cosmological parameters). 
As a common behavior, the NN tends to predict values close to the mean of the parameters range, as doing so artificially reduces the loss function \cite{2024/novaes,2022/perez,2022/lu}. 

Below, we discuss how the neural compressed data are employed to infer the posterior distribution for each cosmological parameter. 

\subsection{Parameter inference}

The SBI approach, rather than making assumptions about the form of the likelihood, relies on simulations to implicitly reconstruct (learn) the likelihood $p(x|\theta)$ of the data \cite{2020/cranmer}. 
Here, we use a Neural Posterior Estimation \cite{2016/papamakarios-murray, 2019/greenberg} method implemented in the {\tt sbi} package\footnote{\href{https://github.com/mackelab/sbi}{https://github.com/mackelab/sbi}} \cite{2020/tejero-sbi} to train an NDE, using a given compressed data $x$, to learning the posterior distributions of the parameters of interest, $\theta$. 
As our NDE, we chose the Gaussian Mixture Density Network\footnote{We also tested the Masked Autoregressive Flow (MAF; \cite{2017/papamakarios}) architecture as the NDE and the performance was similar. } (MDN; \cite{1994/bishop-mdn}) architecture, trained to learn a density $q_\phi(\theta|x)$ that approximates the underlying posterior distribution, $p(\theta|x)$, such that $q_\phi(\theta|x) \approx p(\theta|x)$. 

Again, the training is performed 100 times, each of them excluding the 50 $(x,\theta)$ pairs from one of the cosmologies; the 99 cosmologies employed are split with an 80/20 ratio for training and validation. 
The training is performed by varying the neural network parameters, $\phi$, and maximizing the log-likelihood, $\mathcal{L} = \sum_{i \in T} \log q_\phi(\theta_i|x_i)$, with $i$ running over the training set. 
In other words, it minimizes the Kullback-Leibler divergence between $q$ and $p$ \cite{1951/kullback-leibler}. 
The main hyperparameters of the MDN architecture and the training process are chosen through experimentation. 
We verified that most of these values do not strongly depend on the summary statistics, so that, for all our inferences, we use the {\tt ADAM} optimizer, a number of hidden units of the NN of 20 and batch size of 50 (keep other parameters fixed at the default definitions of {\tt sbi}). 
The learning rate, on the other hand, varies from $10^{-3}$ to $10^{-2}$ depending on the combination of summary statistics. 
Our choice of the learning rate also relies on tests for posterior validation, as we discuss below in this section. 
To help prevent overfitting, we use early stopping, interrupting the training after the validation log-likelihood does not increase after 20 epochs. 

Since the randomness of the initialization of an NDE can lead to different posterior approximations, for the $i$th cosmology removed from the training+validation set, we perform the training 20 times and, among the 20 $q_\phi^i$ estimators, we choose the one leading to the largest validation log-likelihood. 
This way, at the end we have a set of 100 $q_\phi^i$ models, one for each cosmology excluded. 
We note that, when performing parameter inference, for a given combination of summary statistics, over the fiducial set of simulations and over the HSC Y1 data, we choose the 5 $q_\phi^i$ models with greater log-likelihood. 
We sample the posterior distribution inferred by each of them, generating 10000 samples each, which are finally combined in one chain\footnote{We verified that averaging over the median values of each posterior of length 10000, as well as over their $1\sigma$ range, provide very similar result as taking the median and $1\sigma$ from the combined posterior with 50000 samples. } of length 50000 used to derive the parameter constraint.  \vspace{0.1cm}

\paragraph*{\bf Individual \textbf{\textit{versus}} joint posterior estimation.} 
Our analysis has two cosmological parameters of interest, $\Omega_m$ and $S_8$, for which we want to apply our inference pipeline to obtain the best approximated posterior and final parameters constraints, as previously discussed. 
There are two possible ways of training our NDE for this purpose: \textit{1D case}: estimating $q_\phi^{i,p}$ for one parameter, $p=\Omega_m$ and $p=S_8$, at a time, such that both $x$ and $\theta$ are one-dimensional data vectors; and \textit{2D case}: estimating $q_\phi^i$ to approximate the joint posterior of the parameters of interest, with $x$ and $\theta$ being two-dimensional. 
We evaluated both scenarios. 
Given that the former leads to slightly tighter constraints from the fiducial set of simulations, we employ the 1D case for our baseline analysis. 
All tests and results presented here consider the 1D case unless stated otherwise. 
\vspace{0.1cm}

\begin{figure}[t]
\includegraphics[width=\columnwidth]{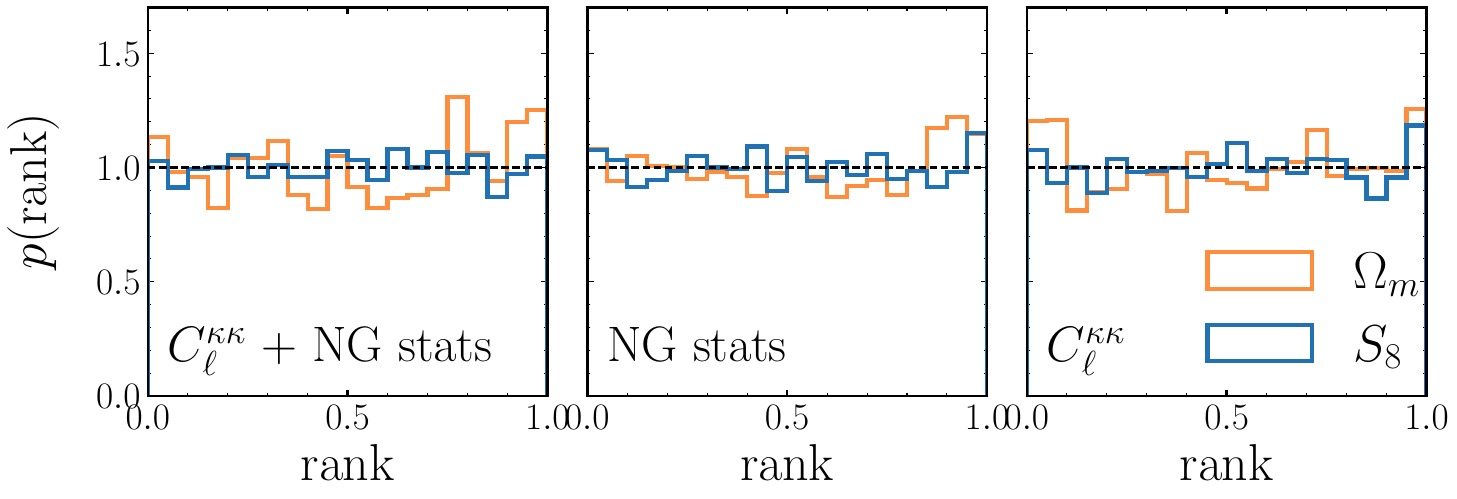}
\caption{\label{fig:rank} Rank distributions constructed using each of the 5000 (50 realizations $\times$ 100 cosmologies) realizations of the cosmology-varied set as one ``observation". 
The histogram is constructed over the 5000 rank values, each of them calculated from 10000 samples drawn for each realization using the corresponding $q_\phi^i$ model. Each panel shows results for $\Omega_m$ (orange) and $S_8$ (blue) using different (combinations of) summary statistics. The rank distribution is expected to be uniform for an accurate (valid) posterior estimate. }
\end{figure}

\paragraph*{\bf Posterior validation.} 
Before applying our method to the data, it is important to ensure that the posterior inference is being performed accurately and robustly. To validate our posteriors, we use the simulation-based calibration, as discussed in Ref. \cite{2020/talts}, and construct the rank statistics for each cosmological parameters of interest. 
Employing a given combination of summary statistics, we define the compressed data, $x_i^n$, from each of the 50 realizations at a time ($n= 1, ..., 50$) as our ``observation". 
We sample the posterior distribution for each realization of the $i$th cosmology using the respective $q_\phi^i$ estimator, for $i = 1, ..., 100$. 
We draw 10000 samples for each $x_i^n$ ``observation" and use them to calculate the rank. 
The rank is computed as the fractional number of samples whose values fall below the true cosmological parameter, $\theta_i$. 
We show in Figure \ref{fig:rank} the rank plot obtained for different combinations of summary statistics. 
For all tests presented, we find quite uniform rank plots, although those for the $S_8$ parameter are more close to flat. 
Rank distributions with $\bigcup$-like shape indicate that the posteriors are overconfident, while the $\bigcap$-like shape indicates that the inferred posteriors are broader than the true one, i.e., they are underconfident and could be improved by more efficient training. 

\section{\label{sec:systematics}Systematics}

We assess the impact of different systematic effects not accounted for in our simulations, quantifying potential bias introduced by each of them, besides evaluating the accuracy of our SBI pipeline when applied outside the training set. 
For this, we perform cosmological inference of fiducial realizations accounting for observational and astrophysical systematics as described below (detailed discussions on the simulations can be found in Ref. \cite{2024/marques-hsc}):
\begin{itemize}[leftmargin=*]
    \item Photometric redshifts: We employ two alternative methods, namely, {\tt Mizuki} and {\tt Frankenz} \cite{2018/tanaka-hsc}, to estimate the redshift, constructing, using each of them, a subset of 100 fiducial realizations. 
    \item Multiplicative bias: Considering the (percent level) uncertainty expected for the multiplicative bias employed for image calibration \cite{2018/mandelbaum}, we introduce a miscalibration of $\Delta m = \pm 0.01$ when generating a subset of 100 fiducial realizations. 
    \item Baryonic feedback: We use the hydrodynamic simulation IllustrisTNG, with and without including baryons, to model the impact of baryonic feedback on the summary statistics. We include its effect directly on the data vectors of the fiducial realizations by rescaling them by the ratio between the summary statistics calculated from the hydrodynamic and gravity-only simulations. 
    \item Intrinsic alignment:  The non-linear tidal alignment model (NLA) quantify the strength of this effect by the coupling parameter $A_{\rm IA}$ \cite{2007/bridle-king}. 
    Given the alignment amplitude estimated for HSC data \cite{2019/hikage-hsc,2020/hamana-hsc}, we choose two different values for this parameter ($\pm 1\sigma$ shift from inferred value), $A_{\rm IA}= 1.18$ and $-0.32$, to contaminate our data vectors. Similarly to the baryonic feedback case, we rescale them by the ratio of the summary statistics estimated from simulations with the chosen $A_{\rm IA}$ and with $A_{\rm IA} = 0$ \cite{2022/harnois-deraps}.
\end{itemize}
For each realization of the fiducial set, taking all four summary statistics as our data vector, we employ the ${\mathcal F}_i$ trained models, for $i = 1, ..., 100$, for data compression and use the mean value (over the 100 compressions) as our ``observation", sampling the posterior distribution for the $S_8$ parameter, as detailed in previous section. 
We follow the exact same procedure for parameter inference from each simulation accounting for a systematic effect.

\begin{figure}[t]
\includegraphics[width=0.9\columnwidth]{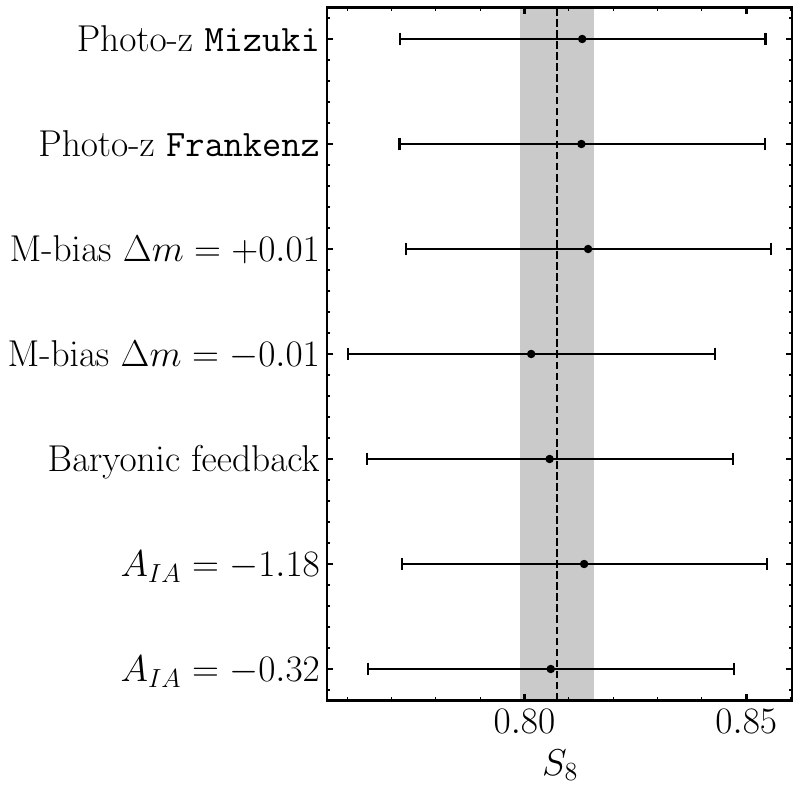}
\caption{\label{fig:systematics} Impact of the different systematic effects on $S_8$ constraints from fiducial simulations. 
The horizontal axis shows the average $S_8$ constraints from data vectors contaminated by systematics.
The vertical axis indicates the individual systematic effect tested. The error bar corresponds to the average 68\% confidence interval of the posteriors obtained for each contaminated data vector. The dashed vertical line represents the average from original fiducial realizations, $S_8 = 0.807 \pm 0.041$, while the gray region illustrates the $\pm 0.2\sigma$ range from the same set. }
\end{figure}

We show in Figure \ref{fig:systematics} the impact on $S_8$ constraints from the fiducial set of simulations due to the contamination by different systematic effects. 
Notice that the vertical dashed line represents the average constraint from the fiducial set, showing a slight deviation from the true value ($0.016\sigma$). 
We find that the bias introduced by systematic effects is no larger than 0.17$\sigma$, obtained for the case of a multiplicative bias miscalibration of  $\Delta m = +0.01$. 
The rather small bias introduced by the evaluated systematic effects confirms the robustness and reliability of our results (see next section). 
In fact, our choices for the combination of all summary statistics, the smoothing scale of 4 arcmin, the exclusion of the highest redshift bin ($z_4$), and the scale cuts on $C_\ell^{\kappa \kappa}$, defining our baseline analysis, are made in such a way to assure the small amplitude of these biases, as we show here\footnote{See in Appendix \ref{appendix:blind} a brief discussion on how we adapt here the blind strategy followed by previous analyses of HSC Y1 data \cite{2024/marques-hsc, 2023/thiele-hsc}.}.
In other words, the small biases we show in Figure \ref{fig:systematics} corroborate the applicability of our pipeline out of the training set, given that none of the systematic effects is considered in the cosmology-varied simulations employed during training processes. 
Finally, it is also worth noticing that no cuts are applied to the NG statistics, meaning that the analyses employing these statistics (individually or combined) use the complete data vectors (all 19 $\nu_t$ bins from each of them). 
The reason for this choice is that using our methodology, unlike the $C_\ell^{\kappa \kappa}$, the NG statistics (combined with $C_\ell^{\kappa \kappa}$; Figure \ref{fig:systematics}) perform better without any cut, and no significant bias is observed.

\section{\label{sec:results}Results and discussions}

In this section, we present our main results and findings analyzing the HSC Y1 data, presenting our cosmological constraints using the combination of all summary statistics, our baseline analysis, and results from other combinations of statistics. 
We also present and discuss some consistency tests, reinforcing the accuracy of our main result. 

\subsection{HSC Y1 analysis}

\begin{figure}
\includegraphics[width=\columnwidth]{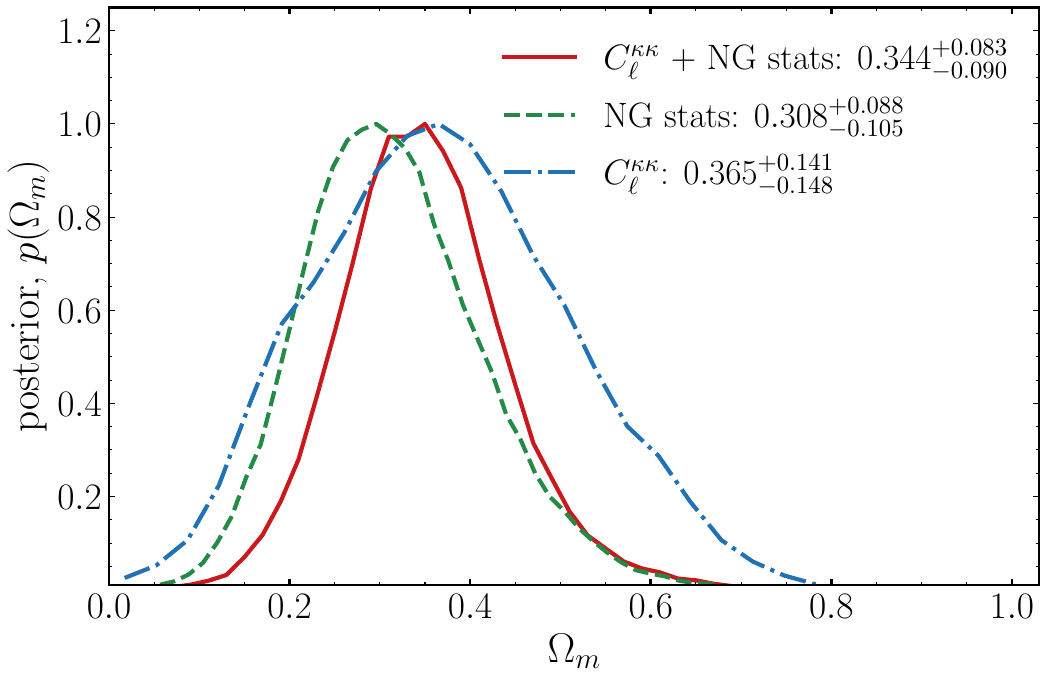}
\includegraphics[width=\columnwidth]{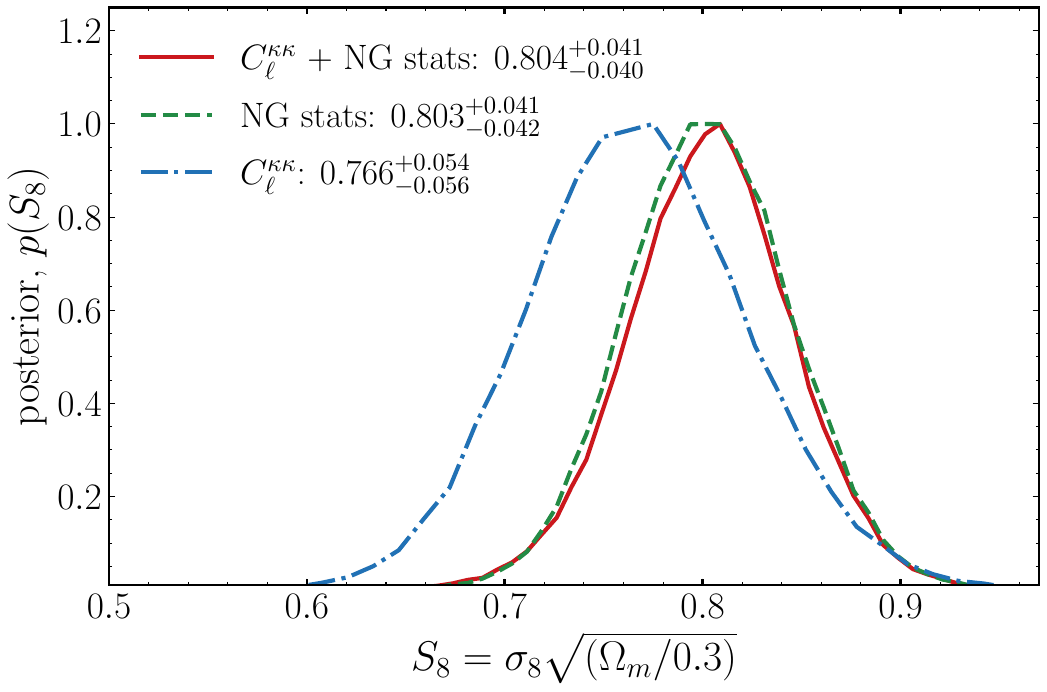}
\caption{\label{fig:1D-posteriors} Cosmological constraints on HSC Y1 data. The $\Omega_m$ and $S_8$ posterior are shown in the top and bottom panels, respectively. Their median and 16 and 84 percentiles are quoted in each panel, showing cosmological constraints using different combinations of summary statistics.  }
\end{figure}

We present in Figure \ref{fig:1D-posteriors} results of our method applied to the HSC Y1 data, showing $\Omega_m$ and $S_8$ posteriors and final constraints obtained using a combination of Gaussian and NG summary statistics, namely, $C_\ell^{\kappa \kappa}$, MFs, PM, and PDF using the analysis described above. 
The constraints obtained from $C_\ell^{\kappa \kappa}$ individually and the NG statistics are also shown for comparison, showing that all three SBI constraints are statistically consistent. 
We find that the combined analysis can improve over the PS constraints by 26\% for $S_8$ and by 40\% for $\Omega_m$, while slightly shifting the posterior median to higher and smaller values, respectively. 
The individual usage of NG statistics seems to prefer higher (smaller) values of $S_8$ ($\Omega_m$) parameter and allows 25\% (30\%) tighter constraints, with respect to PS only. 
It is worth noticing that a similar trend is observed from the {\sc rmse} values shown in Figure \ref{fig:NNpred}. 
Also, the NN predictions for HSC Y1 data, as well as the error bars given by the {\sc rmse} amplitudes, are very consistent ($< 1\sigma$) with the constraints from $\Omega_m$ and $S_8$ posteriors (Figure \ref{fig:1D-posteriors}), in particular, when combining all summary statistics. 
Such consistency not only provides further confirmation of the reliability of our constraints, but also corroborates an expected equivalence between the Bayesian analysis, through posterior estimation, and the frequentist approach, using directly the NN predictions. 

\begin{figure}[t]
\includegraphics[width=\columnwidth]{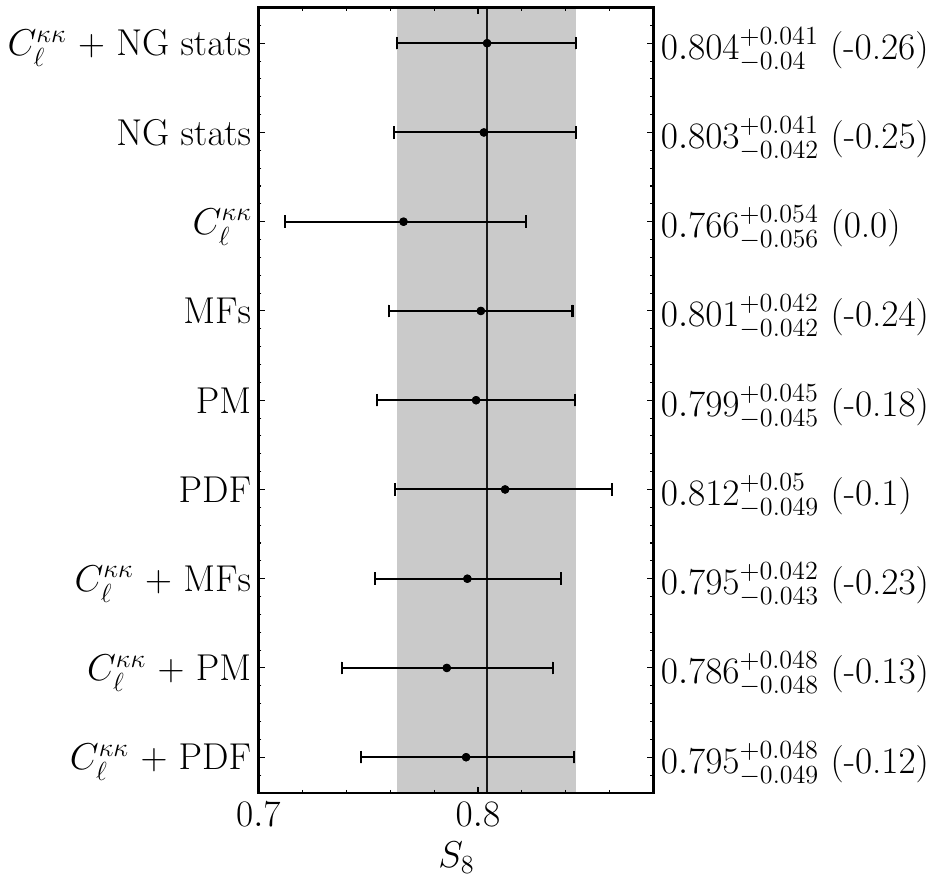}
\caption{\label{fig:all-stats} $S_8$ constraints for HSC Y1 data using different summary statistics and combinations of them, where the error bars correspond to 16 and 84 percentiles. The right vertical axis also indicates, in parentheses, the fractional improvement of each summary statistic for $C_\ell^{\kappa \kappa}$. The dashed vertical line and gray region represent the inference for our baseline analysis ($C_\ell^{\kappa \kappa}$ + NG stats). }
\end{figure}

A summary of the $S_8$ cosmological constraints for the different NG statistics individually and each of them combined with the PS is presented in Figure \ref{fig:all-stats}; results from Figure \ref{fig:1D-posteriors} are also included for comparison. 
Figure \ref{fig:all-stats} shows that all our tests lead to statistically consistent estimates of the $S_8$ parameter, with an apparent preference of $C_\ell^{\kappa \kappa}$ for a smaller value, but still inside the $1\sigma$ confidence interval. 
All the results employing NG statistics only lead to tighter constraints than the PS, following the order of constraining power: MFs $>$ PM $>$ PDF. 
The individual combination of MFs and PDF with the PS yields a tighter error bar.
For PM this is not observed, and its combination with PS yields a slightly wider error compared to PM only, possibly a statistical fluctuation. 
In addition, results from MFs indicate that this NG statistic seems to be primarily responsible for the improvement over the PS-only results, shrinking the $S_8$ error bar by $\sim$24\% in comparison to the PS only constraints. 
It is worth noting that, following an SBI approach, the NG statistics combined with the PS seem to have very similar constraining power as the NG statistics only. 
However, one should point out that our conservative usage of the $C_\ell^{\kappa \kappa}$, with severe scale cuts, could result in the NG statistics capturing most of the information contained in the two-point statistic while retaining almost all the constraining power. 
Therefore, in addition to the improvement of the NG statistics in comparison to the PS only constraints, the robustness against systematics (see Section \ref{sec:systematics}), which allows keeping all $\nu_t$ bins, corroborate the significant contribution of these statistics. 
We emphasize that our conclusions on the relevance of the PS are driven from our analysis constraining only two cosmological parameters. 
Including more parameters might lead to a different constraining power of the PS.

Our SBI analysis leads to constraints that are statistically consistent with previous results derived from HSC Y1 data. 
In terms of two-point information, the official HSC Y1 analyses using the PS and the two-point correlation function find $S_8 = 0.780^{+0.030}_{-0.033}$ \cite{2019/hikage-hsc} and $S_8 = 0.804^{+0.032}_{-0.029}$ \cite{2020/hamana-hsc}, respectively; notice the preference of the PS for a slightly smaller value, similarly to what we find here (see Figure \ref{fig:all-stats}).
Interestingly, constraints obtained by Ref. \cite{2023/liu} indicate that a halo-based theoretical peaks model also prefers slightly smaller values of $S_8$. 
Using a CNN, Ref. \cite{2023/lu-hsc} also provides a $S_8$ constraint quite consistent with our median estimates, besides also pointing to an improvement up to 24\% compared to their PS analysis, similar to our finding including NG statistics.
It is important to emphasize here that differences with respect to our findings are expected since we implement other choices of priors, scale cuts (for PS analyses), and treatment of systematics. 
Also, we employ a completely distinct methodology. 
While analyses from Refs \cite{2019/hikage-hsc,2020/hamana-hsc} are conducted using perturbation theory for modelling the two-point information, and Ref. \cite{2023/liu} considers theoretical models and approximations for the likelihood, we employ an inference pipeline fully based on simulations. 
Also, we use an NDE to learn the posterior distribution of the parameters, while Ref. \cite{2023/lu-hsc}, although also forward modelling the observation, still make assumptions to calculate the likelihood of their data vector (the output from the CNN). 

\subsection{Consistency tests}

\begin{figure}[t]
\includegraphics[width=\columnwidth]{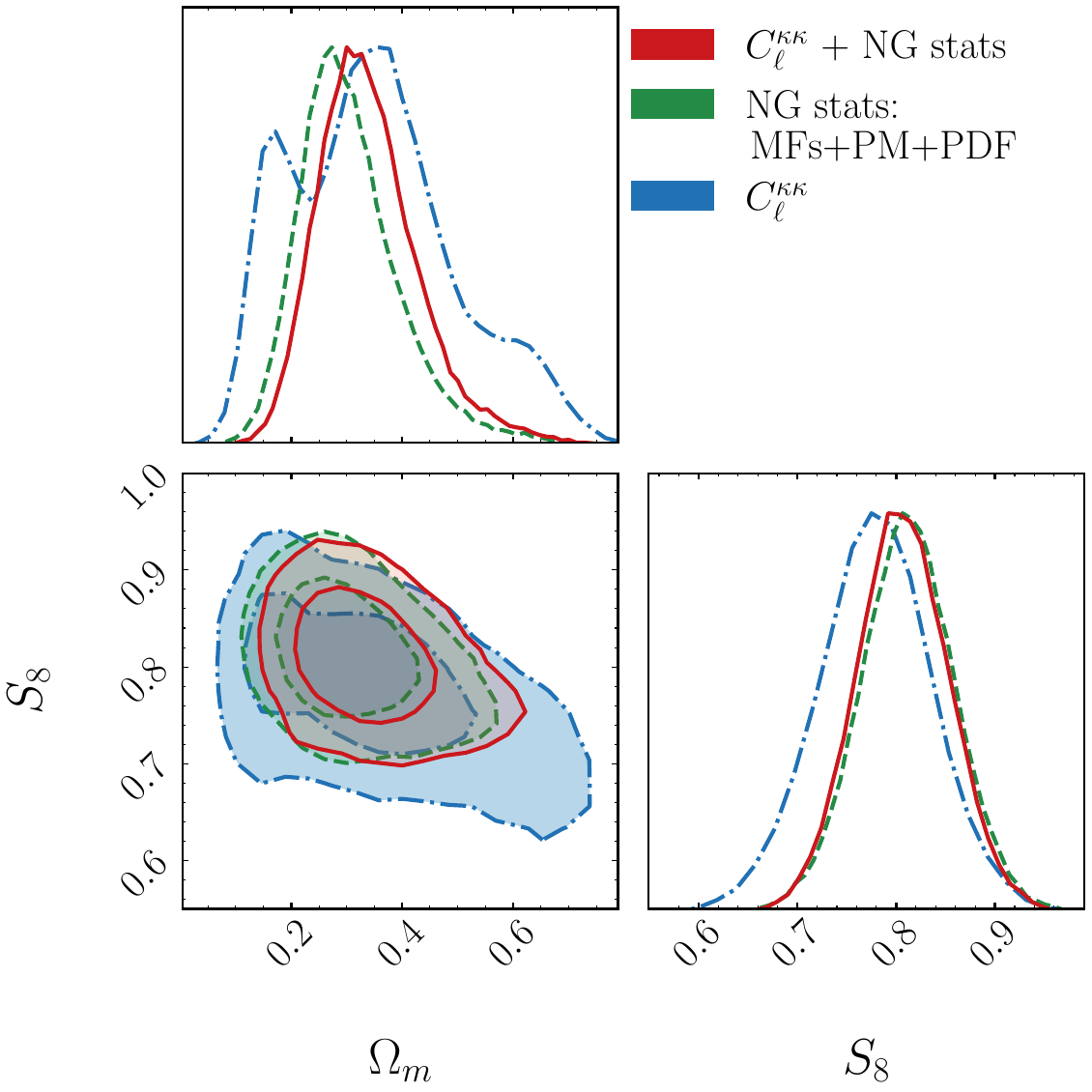}
\caption{\label{fig:2D-posteriors} Posterior distributions (68\% and 95\%) of $\Omega_m$ and $S_8$ with HSC Y1 data using the PS (blue contour), the combination of the three NG statistics (green contour), and the joint analysis combining all four summary statistics (red contour). The results are obtained following the 2D case approach, as discussed in the text (see also Section \ref{sec:sbi} for definition of the 2D case).}
\end{figure}

\begin{figure}[t]
\includegraphics[width=0.85\columnwidth]{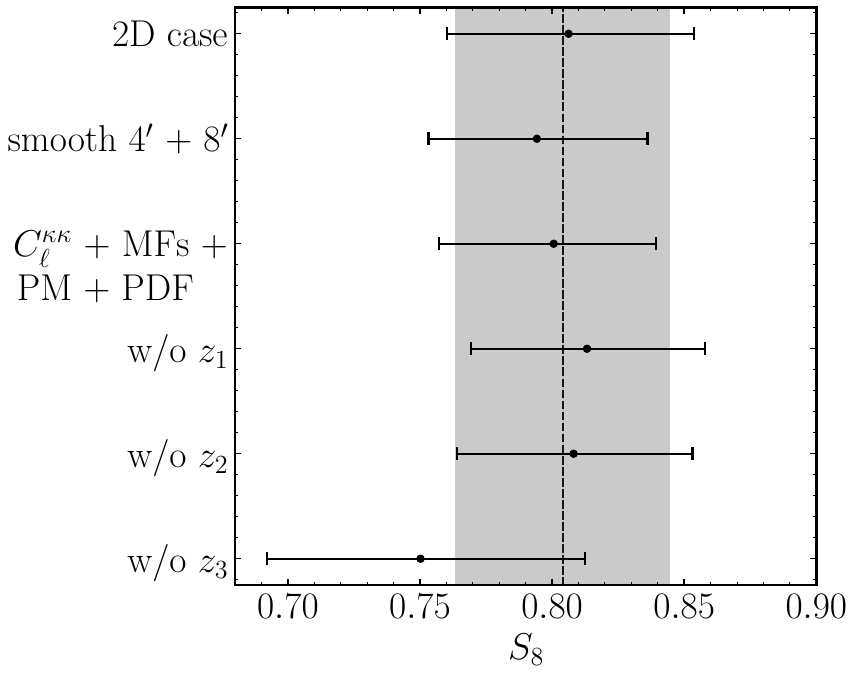}
\caption{\label{fig:consistency} Consistency tests showing the impact on the posterior approximations for $S_8$ due to modifications of our SBI pipeline as employed for the baseline analysis. We show the median of the posterior distributions, with error bars corresponding to the 16 and 84 percentiles. The dashed line and gray region represent the results from the baseline analysis. See text for details on each test performed. }
\end{figure}

To further verify the robustness and reliability of our results, we evaluate our $S_8$ inference against some modifications, one at a time, on our SBI pipeline. 
As we describe in Section \ref{sec:sbi}, our main results are derived through posterior estimation individually for each cosmological parameter of interest ({\it 1D case}). 
Figure \ref{fig:2D-posteriors} shows the $S_8,\Omega_m$ constraints from the joint estimation, the {\it 2D case}.
Combining all summary statistics, we find $\Omega_m = 0.331_{-0.078}^{+0.098}$ and $S_8 = 0.806_{-0.046}^{+0.048}$, highly consistent with results from the {\it 1D case} (Figure \ref{fig:1D-posteriors}). 
The comparison of this 2D estimate of $S_8$ parameter is directly compared to our 1D result in Figure \ref{fig:consistency}, visually confirming the agreement between the median values, obtaining a slightly tighter constraint from the 1D analysis. 

We also evaluate whether the inclusion of compressed information originating from convergence fields with a different smoothing scale would impact our $S_8$ posterior approximation. 
In this case, $x$ is a concatenation of the data vectors obtained for $\theta_s = 4$ arcmin and $\theta_s = 8$ arcmin individually, both resulting from the compression maximizing the information on $S_8$ parameter; $\theta$ is still one-dimensional vector filled with the $S_8$ true values. 
The constraint obtained is consistent with our baseline analysis, as also seen from Figure \ref{fig:consistency} (labelled as ``smooth $4'$ + $8'$"), although we conclude that such additional information does not contribute to a reduction of the uncertainties. 

For all these tests, and our baseline analysis, we apply the data compression over the combination of all four summary statistics combined. 
Our next consistency test evaluates the performance of applying the data compression step to each summary statistic, $C_\ell^{\kappa \kappa}$, MFs, PM, and PDF, individually, and then concatenating the four data vectors into $x$, keeping the same one-dimensional $\theta$. 
Again, we find a median and error bar very consistent with that from our baseline analysis, as shown in Figure \ref{fig:consistency} (results labelled as ``$C_\ell^{\kappa \kappa}$ + MFs + PM + PDF"). 

Finally, we evaluate how the removal of individual tomographic bins from analysis impacts the $S_8$ constraints.
Results are presented in Figure \ref{fig:consistency} and show that the exclusion of $z_1$ and $z_2$ redshift bins do not impact significantly our main result, leading to a slightly increased amplitude of the error bars.
The exclusion of the third redshift bin, $z_3$, on the other hand, has a more significant impact, resulting in a wider posterior with a median shifted to a smaller $S_8$ value, although still consistent within $1\sigma$ with our baseline analysis. 
A similar impact was observed by previous analyses of HSC Y1 using NG statistics \cite{2024/marques-hsc, 2023/thiele-hsc}; in particular, Ref. \cite{2024/cheng-hsc} found, regardless of the summary statistic, a lowering in $S_8$ due to the absence of $z_3$, pointing out to a limitation of $S_8$ constraints by photometric redshift estimation when using HSC. 
For further discussion on this, we refer the reader to Ref. \cite{2024/cheng-hsc}. 
In summary, all tests indicate the robustness of our results, showing consistency within $1\sigma$ for all modifications of the pipeline.

\section{\label{sec:conclusion}Conclusions}

This work presents the first cosmological inference with the HSC Y1 NG statistics using an approach fully based on simulations and no assumptions on the likelihood of the data. 
We forward model a set of Gaussian and NG statistics, namely, the PS, MFs, PM, and PDF, using a suite of N-body simulations tailored to the data. 
We also preprocess our convergence maps (data and simulations) and data vectors by applying smoothing kernels and cuts on scales (for the PS) and redshift bins so that we mitigate possible systematic effects not accounted in the simulations. 
Our SBI pipeline for cosmological inference is implemented in two steps: first, a neural network performs a massive compression of the summary statistics; then, the output (the compressed data) is used to train an NDE to learn the posterior distribution of the parameters of interest. 
The full pipeline is carefully validated on simulations prior to its application to data.

Combining all four summary statistics, we find $S_8 = 0.804_{-0.040}^{+0.041}$ and $\Omega_m = 0.344_{-0.090}^{+0.083}$, which represents an improvement of 26\% and 40\%, respectively, over the PS-only constraints, namely, $S_8 = 0.766_{-0.056}^{+0.054}$ and $\Omega_m = 0.365_{-0.141}^{+0.148}$ (68\% CL). 
Similar constraints are obtained directly from the NN predictions, reinforcing the robustness of our estimates while indicating a consistency between Bayesian and frequentist analyses from the SBI perspective. 
These results are in good agreement (within $1\sigma$) with previous analyses of the same data set, using both Gaussian and NG statistics, as well as with CMB measurements by Planck \cite{2020/aghanim-planck}. 
We also find that the combination of all four summary statistics has very similar constraining power as the NG statistics only (PS absent). 
A possible reason for that might be the severe scale cuts applied to $C_\ell^{\kappa \kappa}$, intended to minimize the impact of systematics. 
This suggests that most of the information in the PS is also captured by the NG statistics, besides the exclusive NG information they are capable of extracting from the convergence fields. 
In particular, using each statistic individually, we also conclude that the information is primarily coming from the MFs, while finding good agreement between the statistics. 
Also, for our SBI approach, no cuts seem to be necessary for the NG statistics, given our choice of $\nu$ range, indicating a certain robustness to systematic effects. 

We verify the robustness of our pipeline against systematic effects.
These effects are the photometric redshift uncertainties, multiplicative bias, baryonic feedback, and intrinsic alignments, not accounted for in the training simulations. 
Such evaluation not only guided choices for the baseline analysis, such as the smoothing scale, redshift bins, and combination of summary statics, but also allowed to confirm the robustness of our final pipeline outside the training set. 
Consistency tests, modifying some of our choices for the baseline analysis, show robust agreement with our constraints, reinforcing the reliability of the results. 

Our study corroborates previous findings on the high constraining power of NG statistics, alone and jointly to standard two-point statistics, for weak lensing analyses. 
We also demonstrate how competitive (and computationally feasible) an SBI approach can be for cosmological inference using as many summary statistics as needed, without making assumptions about the likelihood of the data. 
We show this to be achievable even with a limited number of simulations, a common challenge in weak lensing analyses. 
However, to ensure the minimum impact from unknown systematics, we take actions that result in the loss of information, such as scale cuts and large smoothing scales. 
Future analyses, with improvements in the modelling of systematics, and the use of the latest HSC releases, would increase the constraining power of our methodology. 
Additionally, in the context of the next generation of extragalactic experiments, the straightforward combination of summary statistics from different probes can contribute in exploring additional parameters and alternative models, thereby helping to address the challenges of future surveys.

\begin{acknowledgments}
The authors thank Tilman Tr\"oster and Natalí de Santi for enlightening discussions. 
CPN and EF thanks Instituto Serrapilheira for financial support. 
This work was initiated at the CD3 x Simons Foundation workshop “AI-driven discovery in physics and astrophysics” from January 22-26, 2024, at Kavli IPMU, Kashiwa, Japan. This work was supported by JSPS KAKENHI Grant Number JP23K19064 (JA).
This work was supported by JSPS KAKENHI Grants 23K13095 and 23H00107 (to JL). 
This manuscript has been authored by Fermi Research Alliance, LLC under Contract No. DE-AC02-07CH11359 with the U.S. Department of Energy, Office of Science, Office of High Energy Physics. We acknowledge the use of many public Python packages not cited along the main text: {\tt Numpy}  \citep{2011/numpy}, {\tt Astropy}\footnote{\url{http://www.astropy.org}} a community-developed core Python package for Astronomy~\citep{2018/astropy, 2013/robitaille-astropy}, {\tt Matplotlib} \citep{2007/matplotlib}, {\tt IPython}~\citep{2007/ipython}, {\tt Scipy}~\citep{2020/scipy}, {\tt scikit-learn} \citep{2011/scikit-learn} and {\tt TensorFlow} \citep{2015/tensorflow}. 
This research used computing resources at Kavli IPMU. This research used resources at the National Energy Research Scientific Computing Center (NERSC), a U.S. Department of Energy Office of Science User Facility located at Lawrence Berkeley National Laboratory, operated under Contract No. DE-AC02-05CH11231. 
The Kavli IPMU is supported by the WPI (World Premier International Research Center) Initiative of the MEXT (Japanese Ministry of Education, Culture, Sports, Science and Technology)
\end{acknowledgments}

\appendix

\section{\label{appendix:blind}Blinding procedure}

Given that HSC Y1 official analyses using two-point statistics \cite{2020/hamana-hsc, 2019/hikage-hsc} were already public when studies using NG statistics by Refs \cite{2024/marques-hsc, 2023/thiele-hsc} were conducted, the authors followed a blinding approach as an honor system. 
Here, we employ the same simulations and NG statistics as these studies, in addition to the MFs \cite{2024/armijo-hsc}, and, for this reason, we take some of their finding as our starting points, skipping or modifying some steps of the same blinding procedure, as we describe below:

\begin{itemize}[leftmargin=*]
    \item We employ a suite of simulations tailored to HSC Y1 data to forward model a the summary statistics. Our whole SBI pipeline, data compression and posterior inference, is constructed on these synthetic data vectors and, then, tested and validated taking individual realizations from the fiducial set as ``observation", assuring the recovery of the input cosmology. 
    \item We selected smoothing scales using as ``observation" the data vectors contaminated by systematic effects, so that we minimize their impact. Our choice of redshift bins and scale cuts on the PS follow findings from Refs \cite{2019/marques, 2023/thiele-hsc}. 
    \item Unblinding:
    \begin{enumerate}[leftmargin=0.55cm]
        \item[1º.] B-modes stage: The summary statistics estimated from B-modes of the fiducial set of realizations are compared to those from real data. Refs. \cite{2019/marques, 2023/thiele-hsc, 2024/armijo-hsc} confirmed that both have noise-like behavior, being statistically consistent. For this reason, we skip this step. 
        \item[2º.] PS stage: We apply the SBI pipeline to HSC Y1 data, using the $C_\ell$ only, and compared the results to those previously published by the official analyses by Refs \cite{2020/hamana-hsc, 2019/hikage-hsc}, as well as to results from Refs \cite{2024/marques-hsc, 2023/thiele-hsc}. We find consistency between the results. We note that at this stage \cite{2024/marques-hsc}, based on results from HSC year 3 \cite{2023/dalal-hsc3, 2023/li-hsc3}, decided to exclude the highest redshift bin due to possible problems with redshift calibration. We adopted the same choice since the beginning of our analysis. 
        \item[3º.] NG statistics stage: Finally, we apply our SBI pipeline to the HSC Y1 data, using both the NG statistics only (MFs+PM+PDF) and their combination with the PS, finding good agreement (within $1\sigma$) with the same previous results. 
    \end{enumerate}
    
\end{itemize}


\bibliography{refs}

\end{document}